\documentclass[preprint,12pt]{elsarticle}

\usepackage{amssymb}
\usepackage{graphicx}
\usepackage{subfigure}
\usepackage{dcolumn}
\usepackage{bm}
\usepackage{verbatim} 
\usepackage{mathrsfs}

\journal{Advances in Space Research}

\begin{document}

\begin{frontmatter}


\title{Chang'e 3 lunar mission and upper limit on stochastic background of gravitational wave around the 0.01 Hz band}

\author[AFDL,AMSS,BH]{Wenlin Tang}
\author[AMSS]{Peng Xu}
\author[AFDL]{Songjie Hu}
\author[AFDL]{Jianfeng Cao}
\author[BACIRI]{Peng Dong}
\author[AFDL]{Yanlong Bu}
\author[AFDL,NAO,UCAS]{Lue Chen}
\author[AFDL,NAO]{Songtao Han}
\author[AMSS]{Xuefei Gong}
\author[NAO,UCAS]{Wenxiao Li}
\author[NAO]{Jinsong Ping}
\author[AMSS]{Yun-Kau Lau\corref{cor1}}
\ead{lau@amss.ac.cn}
\author[AFDL]{Geshi Tang\corref{cor2}}
\ead{tanggeshi@bacc.org.cn}

\address[AFDL]{Science and Technology on Aerospace Flight Dynamics Laboratory, Beijing Aerospace Control Center, Beijing, China.}
\address[AMSS]{Institute of Applied Mathematics, Morningside Center of Mathematics and LESC, Institute of Computational Mathematics, Academy of Mathematics and System Science, Chinese Academy of Sciences, Beijing, China.}
\address[BH]{Department of Aerospace Guidance Navigation and Control, School of Astronautics, Beihang University, Beijing, China.}
\address[BACIRI]{Beijing Institute of Aerospace Control Devices, Beijing, China.}
\address[NAO]{National astronomical observatories, Chinese Academy of Sciences, Beijing, China.}
\address[UCAS]{University of Chinese Academy of Sciences, Beijing, China.}
\cortext[cor1]{Corresponding author}
\cortext[cor2]{Corresponding author}

\begin{abstract}
The Doppler tracking data of the Chang'e 3 lunar mission is used to constrain the stochastic background of gravitational wave in cosmology within the 1 mHz to 0.05 Hz frequency band. Our result improves on the upper bound on the energy density of the stochastic background of gravitational wave in the 0.02 Hz to 0.05 Hz band obtained by the Apollo missions, with the improvement reaching almost one order of magnitude at around 0.05 Hz. Detailed noise analysis of the Doppler tracking data is also presented, with the prospect that these noise sources will be mitigated in future Chinese deep space missions. A feasibility study is also undertaken to understand the scientific capability of the Chang'e 4 mission, due to be launched in 2018, in relation to the stochastic gravitational wave background around 0.01 Hz. The study indicates that the upper bound on the energy density may be further improved by another order of magnitude from the Chang'e 3 mission, which will fill the gap in the frequency band from 0.02 Hz to 0.1 Hz in the foreseeable future.

\end{abstract}


\end{frontmatter}



\section{Introduction}\label{Introduction}

Chang'e 3 is an unmanned lunar exploration mission operated by the China National Space Administration. As part of the second phase of the Chinese Lunar Exploration Program, it was landed on the Moon on 14 December 2013, becoming the first spacecraft to soft-land on the Moon since the Soviet Union's Luna 24 in 1976. At present it is located on the Lunar surface at about $44.12^{\circ}$ N, $19.51^{\circ}$ W and -2640 m in elevation(Cao et al., 2014; Ping, 2014). The tracking of the lander by the Chinese deep space network is still ongoing. Every day the lander is tracked continuously for about two to four hours by two ground stations located at Kashi and Jiamusi within China by means of X band radio waves (uplink and downlink at 8.47 GHz). For the Jiamusi station, the two-way Doppler tracking can reach the measurement accuracy of about 0.2mm/s, with sampling time of one second. The high precision Doppler tracking data of Chang'e 3 encodes information concerning the dynamics of the motion of the Moon relative to the Earth and it is worth understanding better whether it is feasible to extract useful science from the data.

As a starting point of our investigation in this direction, the present work aims to understand possible upper bound on the isotropic stochastic background of gravitational waves (SBGWs) imposed by the Chang'e 3 Doppler tracking data. The stochastic background is of cosmological significance as it contains a component of primordial gravitational waves generated during the beginning stage of our Universe(Maggiore, 2000; Sathyaprakash et al., 2009).

The structure of the Chang'e 3 data suggests that the frequency window around 0.01 Hz would be the appropriate window to be looked at. Further, with Chang'e 4 to be launched at around 2018 and the prospect of deep space exploration beyond the Earth-Moon system after Chang'e 4, it is anticipated that more Doppler tracking data with higher precision will be available to the scientific community in future, the present work then also serves the dual purpose of being a pilot study of Doppler tracking data analysis for future Chinese deep space missions. With this prospect in mind, we undertake meticulous noise analysis for the Chang'e 3 Doppler tracking data, in order to understand the prospect of mitigating various noise sources in the Doppler tracking data in future deep space Chinese missions. In addition, a feasibility study is also undertaken to understand the scientific potential of the Chang'e 4 mission in relation to the stochastic gravitational wave background.

Currently, over the band from $10^{-6}$ Hz up to 1 Hz, the Cassini spacecraft(Armstrong et al., 2003) gives the best constraint on the energy density ( $\Omega_{gw}(f)$ ) of the SBGWs from $1.2\times10^{-6}$ Hz up to $10^{-4}$ Hz, the ULYSSES spacecraft(Bertotti et al., 1995) together with the normal modes of the Earth(Coughlin et al., 2014a) give the best constraint on $\Omega_{gw}(f)$ of the SBGWs from $2.3\times10^{-4}$ Hz up to 0.02 Hz, the Apollo missions(Aoyama et al., 2014) gives the best constraint on $\Omega_{gw}(f)$ from 0.02 Hz up to 0.05 Hz, the Earth's seismic data(Coughlin et al., 2014b) gives the best constraint on $\Omega_{gw}(f)$ from 0.05 Hz up to 0.1 Hz, and the Lunar seismic data(Coughlin et al., 2014c) gives the best constraint on $\Omega_{gw}(f)$ from 0.1 Hz up to 1 Hz. See Figure 10 for an illustration of the sensitivity limits obtained by previous missions or other detection methods in different frequency windows. Upon comparison, we find that the constraint on $\Omega_{gw}(f)$ imposed by the Apollo missions from 0.02 Hz up to 0.1 Hz is by far worse than others in the frequency band ranges from $10^{-4}$ Hz to 1 Hz. As the frequency window of Chang'e 3 overlaps with that of the Apollo missions and at the same time the measurement accuracy of the Doppler tracking data of Chang'e 3 is better than that of the Apollo missions, it is not surprising to obtain that the upper bound on $\Omega_{gw}(f)$ imposed by Chang'e 3 improves on that given by the Apollo missions.

The outline of this paper is as follows. Section 2 describes the algorithm for the data analysis and we work out the power spectral density of the noise in the measured Doppler tracking data of Chang'e 3. Section 3 presents the main results concerning the upper bound on the energy density $\Omega_{gw}(f)$ of the SBGWs in the band from 1 mHz to 0.05 Hz by using the Doppler tracking data of Chang'e 3. A detailed noise analysis of the Doppler tracking data of the Chang'e 3 mission is given in Section 4. Section 5 presents the feasibility study of constraining the SBGWs in the frequency band from 1 mHz to 0.1 Hz using the future Doppler tracking data of Chang'e 4. Some brief remarks are then made to conclude this paper in the final section.

\section{Data analysis}\label{Sec-dataandanalysis}

\begin{figure}
\begin{center}
\begin{tabular}{cc}
\subfigure[The observed two-way range rate given at the Jiamusi station for analysis.]{
\label{Fig-omc532023202_56643_RRate_obs:a}
\includegraphics[width=0.5\textwidth]{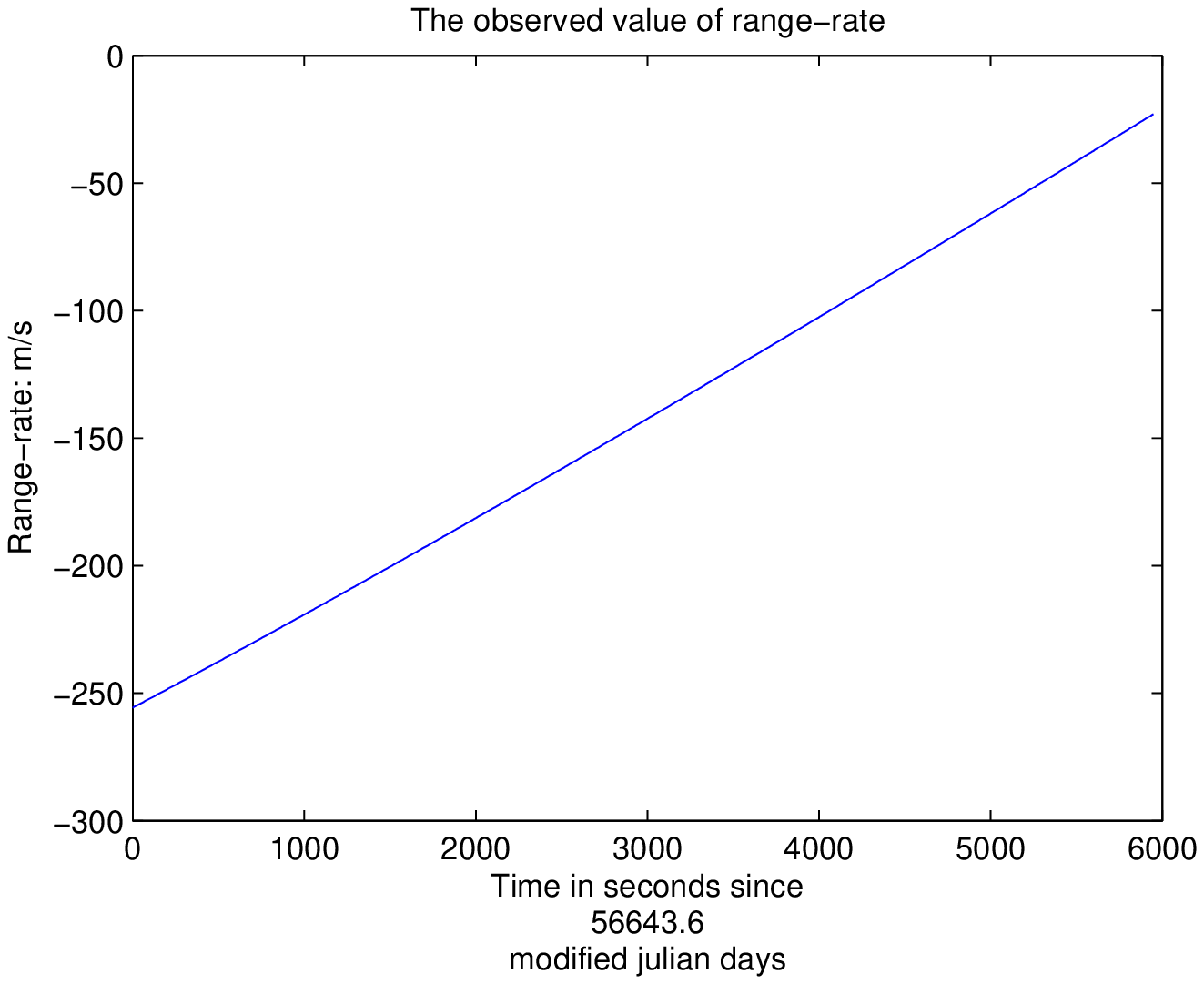}} &
\subfigure[The residual of the corresponding two-way range rate. Its mean is zero and its standard deviation is about 0.2 mm/s.]{
\label{Fig-omc532023202_56643_RRate_obs:b}
\includegraphics[width=0.5\textwidth]{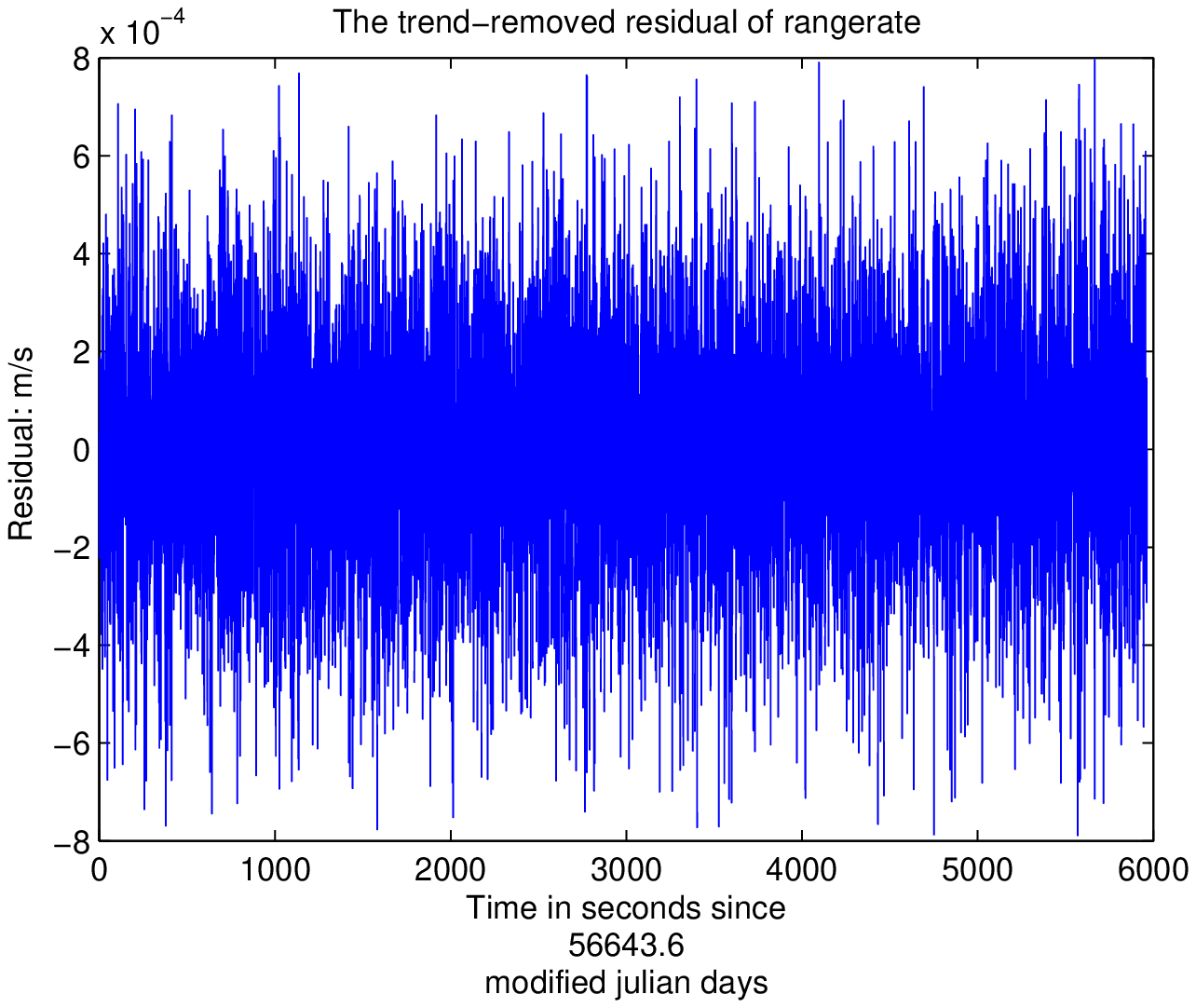}}
\end{tabular}
\caption{The observed two-way range rate for analysis and its residual.}
\label{Fig-omc532023202_56643_RRate_obs}
\end{center}
\end{figure}

The Doppler tracking data used is the two-way range rate data recorded at the Jiamusi station, it is taken from UTC 13:22:25.0 to 15:01:49.0 on 17 December 2013. The time series has sampling time of 1 s and it composes of 6000 data points, with the standard deviation about 0.2 mm/s, as shown in Fig. \ref{Fig-omc532023202_56643_RRate_obs}. The round-trip time between the lander and the station is about 2.65 s. The Sun-Earth-Moon angle was about 180 degrees and therefore the lander was in the solar opposite direction. As we shall see later, the Sun-Earth-Moon angle is an important factor when it comes to the estimate of the tropospheric and ionospheric delay noises in the Doppler tracking.

\begin{figure}[h]
\begin{center}
\begin{tabular}{c}
\includegraphics[width=0.4\textwidth]{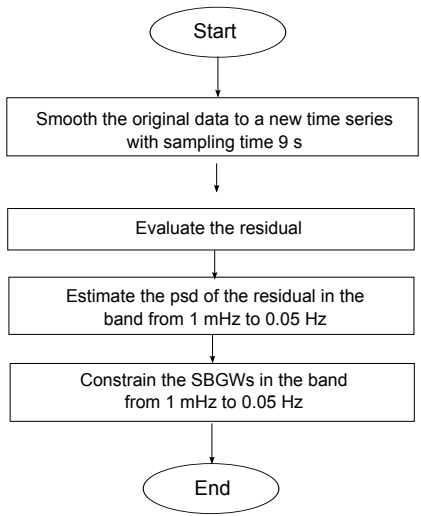}
\end{tabular}
\caption{\label{Fig-CE3_DataProc_FlowChart} The flow chart for the data analysis. }
\end{center}
\end{figure}

The algorithm for data analysis is illustrated in the flow chart displayed in Fig. \ref{Fig-CE3_DataProc_FlowChart}. In this work, we will focus on constraining the SBGWs in the frequency band from 1 mHz to 0.05 Hz using the Doppler tacking data of Chang'e 3. According to the Nyquist criterion, the data used to constrain the SBGWs must have a sampling interval smaller than 10 s. Further, the random fluctuations in a time series can be suppressed by smoothing it to a new time series with longer sampling interval. To this end, we smooth the Doppler tracking data to generate a new time series with sampling time 9 s. Then we evaluate the residual of the new time series and estimate its power spectral density in the band from 1 mHz to 0.05 Hz. Finally, we give the upper bound on the energy density $\Omega_{gw}$ of the SBGWs in the frequency band from 1 mHz to 0.05 Hz.

In the Chang'e 3 mission, the radio signal is transmitted from the station at time $t_1$ to the lander at time $t_2$ and then reflected from the lander at time $t_2$ to the station at time $t_3$. All dates $t_1$, $t_2$ and $t_3$ are in Universal Time Coordinated (UTC). For the Doppler tracking, one observation consists in $t_3$, the date of the radio signal arrived at the station in Universal Time Coordinated (UTC), and $v_o$, the two-way range rate, which is obtained by $v_o=c\cdot \triangle f_o$, where $\triangle f_o$ is the relative variation of the frequency of the radio signal. The other data supplied are the frequency of the radio signal, the signal/noise rate, and the temperature, pressure and humidity of the atmosphere.

Denote by $\{v(t_i)\}_{i=1}^n$ the range rate data, where n is its length. The range rate is smoothed to generate a new time series $\{\bar{v}(s_i)\}_{i=1}^m$ with sampling time 9 s, where $m$ is the largest integer smaller than $n/9$ and $s_i = t_{1+9(i-1)}$. Denote by $\{\epsilon(s_i)\}_{i=1}^m$ the residuals in the time series $\{\bar{v}(s_i)\}_{i=1}^m$ defined by
\begin{equation}\label{Eq-Residual_9sInterval}
\epsilon(s_i) \equiv \bar{v}(s_i) - v_c(s_i), \quad i = 1,2,\cdots,m,
\end{equation}
where $v_c(t)$ is the theoretical value of the two-way range rate at epoch $t$, which is evaluated as
\begin{equation}\label{Calculated_TwoWayRangeRate}
v_c(t) = \frac{\rho^{2w}(t) - \rho^{2w}(t-\triangle t)}{\triangle t}
\end{equation}
from the two-way range $\rho^{2w}$ by difference, where in our calculations, $\triangle t = 1$s.

\begin{figure}[h]
\begin{center}
\begin{tabular}{c}
\includegraphics[width=0.5\textwidth]{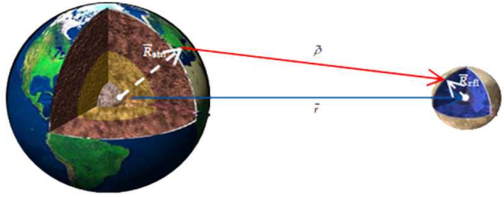}
\end{tabular}
\caption{\label{Fig-ChangE3measure} The Lunar Radio Ranging in the Chang'e 3 mission. It measures the distance between the station on the Earth and the lander on the Moon. This measurement principle is the same as that of the Lunar Laser Ranging, which measures the distance between the station on the Earth and the reflectors on the Moon (Williams et al., 2009).}
\end{center}
\end{figure}

The calculation of the two-way range in Chang'e 3 is introduced as follows. We show in Fig. \ref{Fig-ChangE3measure} the schematic diagram of the lunar radio ranging of Chang'e 3. Compare with the lunar laser ranging (LLR) measurement(Chapront et al., 2006), we find that the role of the lander is similar as that of the reflectors in the LLR. Thus we may follow the same procedure(Chapront et al., 2006) in calculating the round-trip time of light in LLR to calculate the two-way range in Chang'e 3.

The theoretical value of the two-way range can be evaluated by $\rho^{2w}=c \cdot \triangle t$ from the duration $\triangle t$ of the round trip travel of the radio signal in atomic time (TAI, Temps Atomique International), where $c$ is the speed of light. To model the duration $\triangle t$, we need to take into account the relativistic curvature of the signal, and the influence of the troposphere and ionosphere. The computation of the duration $\triangle t$ should be given in the frame of General Relativity theory, and will depend on the barycentric positions of $T$, $L$ and $S$, respectively the center of the mass of the Earth, of the Moon and of the Sun. Thus in the calculation all coordinated are in the celestial barycentric reference system. The theoretical value of $\triangle t$ is given by(Chapront et al., 2006)
\begin{equation}\label{Calculated_Duration}
 \triangle t = [t_3 - \triangle T_1(t_3)] - [t_1 - \triangle T_1(t_1)],
\end{equation}
which is the same as that used in the calculation of the round trip time of the light, where $\triangle T_1$ is the relativistic correction on the time scale, it transforms the time in Barycentric Dynamical Time (TDB) to that in TAI, whose calculation can see the reference(Soffel et al., 2013). However, since the observation data contains only the date $t_3$, $t_1$ (or $\triangle t$) should be given by iteration from $t_3$ and the ephemerids of the Earth, the Moon and the Sun, such as
\begin{eqnarray}
  t_3 &=& t_2 + \frac{1}{c} |\textbf{BR}(t_2)-\textbf{BO}(t_3)|+\triangle T_{grav}+\triangle T_{trop}+\triangle T_{iono},\\
  t_2 &=& t_1 + \frac{1}{c} |\textbf{BR}(t_2)-\textbf{BO}(t_1)|+\triangle T_{grav}+\triangle T_{trop}+\triangle T_{iono},
\end{eqnarray}
where $B$ is the barycenter of the solar system, $\textbf{BR}$ and $\textbf{BO}$ are respectively the coordinates of the lander and the station with respect to $B$, $\triangle T_{grav}$ is the time contribution due to the gravitational curvature of the signal, $\triangle T_{trop}$ is the atmospheric delay and the $\triangle T_{iono}$ is the ionospheric delay.

The vectors $\textbf{BR}$ and $\textbf{BO}$ are given by(Chapront et al., 2006)
\begin{equation}
  \textbf{BR}(t) = \textbf{BG}(t)+\frac{m_T}{m_T+m_L}\textbf{TL}(t)+\textbf{LR}(t)
\end{equation}
and
\begin{equation}
  \textbf{BO}(t) = \textbf{BG}(t)-\frac{m_L}{m_T+m_L}\textbf{TL}(t)+\textbf{TO}(t),
\end{equation}
where $m_T$ and $m_L$ are respectively the Earth and Moon masses, $\textbf{BG}(t)$ is the coordinates of the Earth-Moon barycenter $G$ with respect to $B$, $\textbf{TL}(t)$ is the position vector from the center of mass of the Earth to that of the Moon, $\textbf{LR}(t)$ is the position vector from the center of mass of the Moon to the lander, and $\textbf{TO}(t)$ is the position vector from the center of mass of the Earth to the station. In our paper, the vectors $\textbf{BG}(t)$ and $\textbf{TL}(t)$ are provided by the JPL planetary ephemeris DE421. The vector $\textbf{LR}$ is dependent of the lander coordinates, which are $(1173217.870, -416319.429, 1208153.007) (m)$(Wagner et al., 2014) in a selenocentric frame defined by the principal axes of inertia of the Moon. The coordinates of the lander should be transformed from the selenocentric frame to the celestial barycentric reference system(Chapront et al., 2006). The vector $\textbf{TO}$ are primarily defined in the International Terrestrial Reference Frame (ITRF) and are subject to various corrections due to the Earth deformations: terrestrial and oceanic tides and pressure anomaly. The transformation from the ITRF to the celestial barycentric reference system involves the Earth rotation parameters, the precession, the nutation, the obliquity $\epsilon$ in J2000.0 and the arc $\phi$ separating the inertial equinox and the origin of the right ascensions on the equator of J2000.0(Chapront et al., 2006).

The correction $\triangle T_{grav}$ due to the gravitations of the Moon, Earth and Sun can be given according to the one-body light time equation(Moyer, 2003). The tropospheric delay $\triangle T_{trop}$ and ionospheric delay $\triangle T_{iono}$ on the radio signal are evaluated according to the tropospheric model and ionospheric model recommended by IERS convention No.36(Petit et al., 2010). To calculate the ionospheric delay $\triangle T_{iono}$, the total electron content (TEC) used is provided by International GNSS Service (IGS) associate analysis centers (see http://cddis.nasa.gov).

\begin{figure}[h]
\begin{center}
\begin{tabular}{c}
\includegraphics[width=0.42\textwidth]{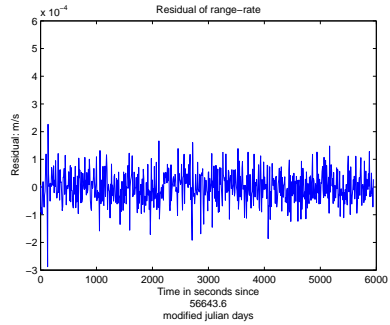}
\end{tabular}
\caption{\label{Fig-omc532023202_56643_vel_Chebfit_9sInterval} The residual of the time series $\{\bar{v}(s_i)\}_{i=1}^m$ used to constrain the SBGWs in the frequency band from 1 mHz to 0.05 Hz. Its mean is zero and its root mean square is about $5.92\times 10^{-5}$ m/s.}
\end{center}
\end{figure}

When we obtain the theoretical value of the two-way range, the theoretical value of the two-way range rate can be obtained from the Eq. (\ref{Calculated_TwoWayRangeRate}). The final residual is shown in Fig. \ref{Fig-omc532023202_56643_vel_Chebfit_9sInterval}. Its root mean square is about $5.92\times 10^{-5}$ m/s, which is consistent with the measurement accuracy of the original two-way range rate data in Chang'e 3.

\begin{figure}[h]
\begin{center}
\begin{tabular}{c}
\includegraphics[width=0.42\textwidth]{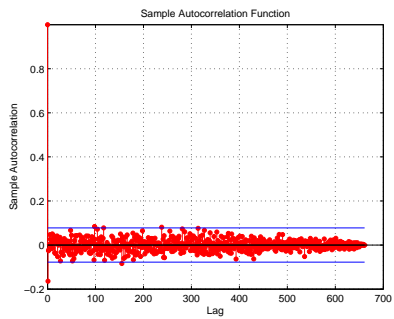}
\end{tabular}
\caption{\label{Fig-omc532023202_56643_vel_autocorr_II_9sInterval} The sample autocorrelation coefficients of the residual. The lower and upper bounds( the two blue lines) in the figure represent the 2$\sigma$ bounds of the estimated autocorrelation coefficients of the residual. The residual of the range rate data measured by the Deep Space Station at Jiamusi. The data is obtained at modified Julian date 56643.}
\end{center}
\end{figure}

The power spectral density of the residual is estimated as follows. Our approach is to fit the residual by a standard autoregressive moving average (ARMV) process(Peter et al., 1991). Then estimate the power spectral density of the residual by that of the fitted process. The sample autocorrelation coefficients of the residual are evaluated as shown in Fig. \ref{Fig-omc532023202_56643_vel_autocorr_II_9sInterval}. From this figure, we find that for all nonzero time lags only $\rho(1) = -0.1639$ exceed the bounds given by the two blue lines in the figure, where the two lines represent the 2$\sigma$ uncertainties if we estimate the autocorrelation coefficients of the residual by the sample autocorrelation coefficients. Here $\sigma$ is about 0.03. Thus we may fit the residual with the standard moving average processing MV(1) model(Peter et al., 1991):
\begin{equation}\label{MV1Model}
X_t = Z_t + \theta Z_{t - 1},
\end{equation}
where $\theta$ is a constant, and $Z_t$ is a Gaussian process with mean zero and variance $\sigma_z^2$. The values of $\theta$ and $\sigma_z$ can be estimated from the root mean square and the sample correlation coefficient $\rho(1)$ of the residual as
\begin{eqnarray}
  \theta &=& -0.1639, \label{MV1ThetaEstVal} \\
  \sigma_z &=& 5.84 \times 10^{-5}. \label{MV1SigmaEstVal}
\end{eqnarray}
Therefore the noise in the new range rate data is estimated by the MV(1) process (\ref{MV1Model}), where the parameter $\theta$ and the root mean square $\sigma_z$ of the Gauss process $Z_t$ are given by Eq. (\ref{MV1ThetaEstVal}) and Eq. (\ref{MV1SigmaEstVal}) respectively.

\begin{figure}[h]
\begin{center}
\begin{tabular}{c}
\includegraphics[width=0.42\textwidth]{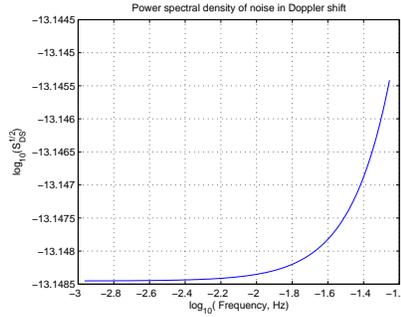}
\end{tabular}
\caption{\label{Fig-omc532023202_56643_DS_PSD_9sInterval} The power spectral density of the noise in Doppler shift. The data is obtained at modified Julian date 56643 and at the deep space station at Jiamusi.}
\end{center}
\end{figure}

For the MV(1) process given in Eq. (\ref{MV1Model}), from its spectral density(Peter et al., 1991)
\begin{equation}\label{PsdOfMV1Process}
S(f) = \frac{\sigma_z^2}{2\pi}[1 + 2\theta \cos(2\pi f) + \theta^2],
\end{equation}
we may estimate the power spectral density of the noise in the new range rate in the band from 0.1 mHz to 0.05 Hz to be
\begin{equation}\label{PsdOfNoiseInRR}
S_{vel}(f) = 5.528\times 10^{-10}\cdot\left[1 - 0.3278\cos(2\pi f) +  0.0269\right].
\end{equation}
Since the Doppler shift $\triangle f / f$ is obtained from the two-way range rate $v^{2w}$ by $\triangle f / f = v^{2w}/c $, the power spectral density of the noise in the Doppler shift in the band from 0.1 mHz to 0.05 Hz is estimated from Eq. (\ref{PsdOfMV1Process}) as
\begin{equation}\label{PsdOfNoiseInDS}
S_{DS}(f) = 6.039\times 10^{-27}\cdot\left[1 - 0.3278\cos(2\pi f) +  0.0269\right],
\end{equation}
and it is shown in Fig. \ref{Fig-omc532023202_56643_DS_PSD_9sInterval}.

\section{Stochastic background of gravitational waves' upper bound in the 0.02 to 0.05 Hz band}\label{sec_SBGWUB}

\begin{figure}[h]
\begin{center}
\begin{tabular}{c}
\includegraphics[width=0.42\textwidth]{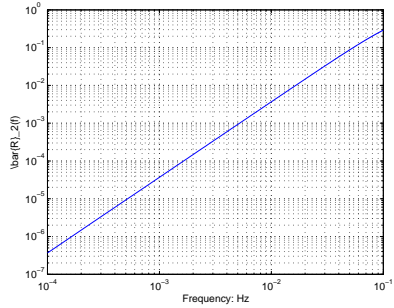}
\end{tabular}
\caption{\label{Fig-TransferFunShow1e-421e-1} Transfer function of Doppler shift w.r.t SBGW when $T_2 = 2.65 $ s.}
\end{center}
\end{figure}

Using spacecraft Doppler tracking to detect gravitational waves was first proposed by Estabrook and Wahlquist in 1975(Estabrook et al., 1975). Let $S^{gw}_{y2}(f)$ be the power spectrum of the two-way fractional Doppler fluctuations generated by the isotropic gravitational wave background. It is related to the power spectrum $S_h(f)$ of the stochastic background of gravitational waves by $S^{gw}_{y_2}(f) = \bar{R}_2(f)S_h(f)$, where $\bar{R}_2(f)$ is the transfer function. For each Fourier component of the SBGWs, the transfer function is given as(Estabrook et al., 1975)
\begin{eqnarray}\label{RespFuncFreqSBGW_EandW}
  \bar{R}_2(f) &=& 1 - \frac{1}{3}\cos(2\pi f T_2) -\frac{(3 + \cos(2\pi f T_2))}{(\pi f T_2)^2}\nonumber\\
   &&+ \frac{2\sin(2\pi f T_2) }{(\pi f T_2)^3},
\end{eqnarray}
where $T_2$ is the round-trip time of the radio signal transmitted from the station to the lander and then back to the station. Input the mean round-trip time $T_2 = 2.65$ s into the above transfer function and the result is shown in Fig. \ref{Fig-TransferFunShow1e-421e-1}.

The spectrum of the isotropic SBGWs can be characterized by its dimensionless energy density $\Omega_{gw}(f)$, or the characteristic rms strain $S_h(f)$ of the wave, which are defined respectively by(Maggiore, 2000; Armstrong et al., 2003)
\begin{equation}\label{rms_strain}
  S_h(f) = \frac{S^{gw}_{y_2}(f)}{\bar{R}_2(f)},
\end{equation}
and
\begin{equation}\label{ENergydensityOmega}
\Omega_{gw}(f) = \frac{8\pi^2 f^3}{3H_0^2}S_h(f),
\end{equation}
where the $H_0$ denotes the Hubble constant.

\begin{figure}[h]
\begin{center}
\begin{tabular}{c}
\includegraphics[width=0.42\textwidth]{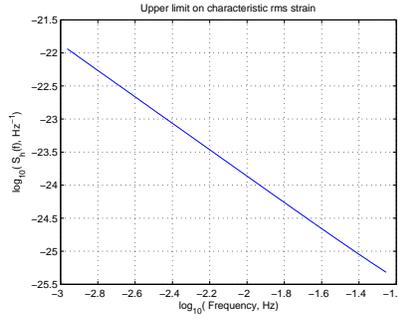}
\end{tabular}
\caption{\label{Fig-omc532023202_56643_UBonSh_9sInterval} The upper bound on the characteristic rms strain $S_h(f)$ of the stochastic background of gravitational wave in the frequency band from 1 mHz to 0.05 Hz.}
\end{center}
\end{figure}

\begin{figure}[h]
\begin{center}
\begin{tabular}{c}
\includegraphics[width=0.42\textwidth]{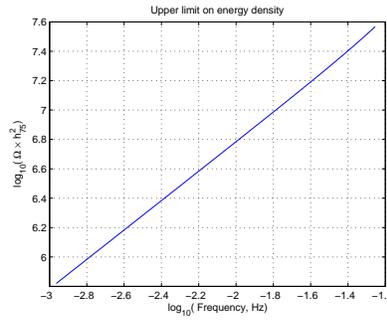}
\end{tabular}
\caption{\label{Fig-omc532023202_56643_UBonOmega_9sInterval} The upper bound on the energy density $\Omega_{gw}$ of the stochastic background of gravitational wave in the frequency band from 1 mHz to 0.05 Hz.}
\end{center}
\end{figure}

At this level of measurement precision of the Doppler tracking data of the Chang'e 3 mission, the fluctuation power generated by the stochastic background of gravitational waves is expected to be submerged under the observed fluctuation power, namely,
\begin{equation}\label{UpperBoundOnS_gw}
 S^{gw}_{y2}(f) \leq S_{DS}(f).
\end{equation}
From Eqs. (\ref{rms_strain}) and (\ref{ENergydensityOmega}), this implies that
\begin{equation}\label{UpperBoundOnS_h}
S_h(f) \leq \frac{S_{DS}(f)}{\bar{R}_2(f)}
\end{equation}
and
\begin{equation}\label{UpperBoundOnOmega_gw}
\Omega_{gw}(f) \leq \frac{8\pi^2 f^3}{3H_0^2}\cdot\frac{S_{DS}(f)}{\bar{R}_2(f)}.
\end{equation}
From the power spectral density $S_{DS}(f)$ of the noise in the Doppler shift given in Eq. (\ref{PsdOfNoiseInDS}), the upper bounds on the characteristic rms strain $S_h(f)$ and the dimensionless energy density $\Omega_{gw}(f)$ are worked out and shown respectively in Fig. \ref{Fig-omc532023202_56643_UBonSh_9sInterval} and Fig. \ref{Fig-omc532023202_56643_UBonOmega_9sInterval}. For the dimensionless energy density $\Omega_{gw}(f)$, the results give the upper bound on the $\Omega_{gw}$ ranging from $6.04 \times 10^5\cdot h_{75}^{-2}$ at 1 mHz to $ 3.24\times 10^7\cdot h_{75}^{-2}$ at 0.05Hz, where $h_{75}$ is the Hubble constant in units of $75 km\cdot s^{-1}\cdot Mpc^{-1}$.

\begin{figure}[h]
\begin{center}
\begin{tabular}{c}
\includegraphics[width=0.42\textwidth]{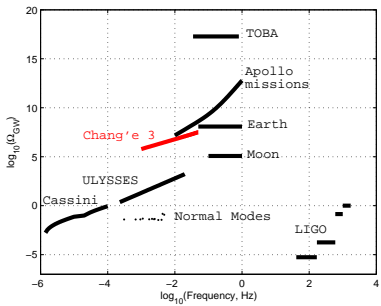}
\end{tabular}
\caption{\label{Fig-omc532023202_56643_UBonOmega_Comparing_I_9sInterval}Summary of the upper bounds on the energy density of SBGWs from $10^{-6}$Hz to $10^4$ Hz band. It includes the results obtained from the Cassini spacecraft(Armstrong et al., 2003), the ULYSSES spacecraft(Bertotti et al., 1995), the normal modes of the Earth(Coughlin et al., 2014a), the Apollo missions(Aoyama et al., 2014), the Earth's seismic data(Coughlin et al., 2014b), the Lunar seismic data(Coughlin et al., 2014c), the torsion-bar antenna(Shoda et al., 2013), the LIGO mission(Aasi et al., 2014)(corresponds to four lines) and the Chang'e 3 mission (red line). }
\end{center}
\end{figure}

Comparing (see Fig. \ref{Fig-omc532023202_56643_UBonOmega_Comparing_I_9sInterval}) the upper bound of the energy density $\Omega_{gw}$ obtained from Chang'e 3 with those given by other missions in the low frequency band from 0.1 mHz to 1 Hz, we find that Chang'e 3 gives the best upper bound in the frequency band from 0.02 to 0.05 Hz. This improves the results given by the Apollo missions(Aoyama et al., 2014), with almost one order of improvement at around 0.05 Hz. Further, at about 0.05 Hz, our result is slightly better than that given by the Earth's seismic data(Coughlin et al., 2014b).

\section{Noise analysis}

In this section we will carry out detailed noise analysis of Chang'e 3. It should be remarked that in the Chang'e 3 mission, the Doppler tracking data was originally used for the determination of the position of the lander on the lunar surface. Thus no detailed calibrations on individual noise source have been made. Only a total noise budget for the Doppler tracking data was considered. What we do is to estimate these individual noise sources indirectly through some auxiliary data, together with the characteristics of the noises and some previous noise analysis of the Doppler tracking data in other missions, such as the Cassini mission. The expectation is that the detailed noise analysis will help us understand the prospect of mitigating various noise sources in the Doppler tracking data in the upcoming Chang'e 4 mission and future Chinese deep space missions.

The main noises in the Doppler tracking in Chang'e 3 contain the propagation noises and the instrumental noises, as listed in Table \ref{Tab-CE3_Noise_Analysis}. These noise terms include the tropospheric noise $y^{trop}$ and the ionospheric noise $y^{iono}$ due to the phase scintillation when the radio signal propagates through the neutral atmosphere and the ionosphere; the antenna noise $y^{ant}$ due to the antenna mechanical motion; the clock noise $y^{FTS}(t)$ due to the instability of the frequency standard; the unmodeled mechanical motion of the lander $y^{ldr}$; the transponder noise $y^{trans}$; the thermal noise $y^{ther}$ in the receiver due to the finite signal-to-noise ratio on the downlink and the ground electrical noise $y^{groundelec}$. These noises enter into the two-way fractional Doppler $y_2(t)$ in a way given by
\begin{eqnarray}\label{CE-3NoiseTransfer}
  y_2(t) &=& y_2^{gw}(t) + y^{trop}(t) + y^{trop}(t - T_2) + y^{iono}(t) + y^{iono}(t - T_2)\nonumber\\
  && + y^{FTS}(t) - y^{FTS}(t - T_2) + y^{ant}(t) + y^{ant}(t - T_2) + y^{ldr}\left(t - \frac{T_2}{2} \right) \nonumber \\
  && + y^{trans}\left(t - \frac{T_2}{2} \right) + y^{ther}(t) + y^{groundelec}(t).
\end{eqnarray}
For any noise, its statistics can be given by the Allan deviation $\sigma_y(\tau)$(Barnes et al., 1971), where $\tau$ is the integration time. The Allan deviation may be evaluated from the spectrum $S_y(f)$ of the noise by(Barnes et al., 1971) $\sigma_y^2(\tau)=\int_{-\infty}^{\infty}df 2S_y(f)\sin^4(\pi f\tau)/(\pi f\tau)^2$.

\begin{table}[h]
\begin{center}
\begin{tabular}{|l|l|}
  \hline
  Noise              & Allan deviation $\sigma_y$ ($\tau = 9$s) \\
  \cline{1-2}
  Clock              & Estimated to be smaller than  $1.0\times10^{-13}$ \\
   \cline{1-2}
  Thermal            & Estimated to be smaller than  $5.5\times10^{-15}$ \\
   \cline{1-2}
  Troposphere        & Estimated to be smaller than $5.0\times10^{-14}$ \\
   \cline{1-2}
  Ionosphere         & Estimated to be smaller than $9.1\times10^{-15}$\\
   \cline{1-2}
  Lander             & Estimated to be smaller than $1.0\times10^{-16}$ \\
  \cline{1-2}
  Transponder        & Estimated to be smaller than $1.0\times10^{-13}$ \\
   \cline{1-2}
  Antenna            & No explicitly positive correlation exists in the  \\
                     & residual, thus expected to be smaller than\\
                     & the noise in the Doppler tracking \\
   \cline{1-2}
  Ground electronics & Estimated not to contribute significantly to \\
                     &  the noise in the Doppler tracking  \\
  \hline
\end{tabular}
\caption{Main noise sources and the Allan deviations of their contributions to the residual of the Doppler tracking data of the Chang'e 3 mission. }
\label{Tab-CE3_Noise_Analysis}
\end{center}
\end{table}

In this work, we will concentrate on the Doppler noise spectra in the frequency band from 1 mHz to 0.1 Hz. Within this frequency band, the instrumental noises are dominated by the noise in the clock, the thermal noise in the receiver, and the antenna mechanical noise(Tinto, 2002; Asmar et al., 2005).

The \textit{clock noise} $y^{FTS}$ is fundamental to the radio observation. It enters into the Doppler shift by $y^{FTS}(t) - y^{FTS}(t-T_2)$. In Chang'e 3, the clock used is the hydrogen masers, its stability can be better than $5\times 10^{-14}$ when $\tau = 9$s. Thus the Allan deviation of the clock noise in the Doppler shift is at most $1.0\times 10^{-13}$. It is smaller than that of the noise of the Doppler shift, because from the power spectral density $S_{DS}(f)$ given in Eq.(\ref{PsdOfNoiseInDS}), the Allan deviation of the noise in the Doppler shift is evaluated to be close to $2.13\times 10^{-13}$ when the parameter $\tau$ is 9 s.

The \textit{thermal noise} is white in phase, which is determined essentially by the finite effective temperature of the receiver and the finite intensity of the signal. The Allan deviation for white phase noise associated with the finite signal-to-noise ratio (SNR) thermal noise component is given by $\sigma_{y}(\tau) \approx \sqrt{3BS_{\phi}}/(2\pi f_0 \tau)$, where $B$ is the bandwidth of the phase detector, $f_0$ is the frequency of the radio signal, and $S_{\phi}$ is the one-sided phase noise spectral density, which is approximated by $1/$(SNR in a 1-Hz bandwidth)(Barnes et al., 1971). For the Chang'e 3 mission, using the X-band observation at the Jiamusi station, its signal-to-noise ratio in the 1 Hz band is larger than 59 dB. Thus its Allan deviation is smaller than $5.5\times 10^{-15}$ when $\tau$ is 9s. According to Eq. (\ref{CE-3NoiseTransfer}), the Allan deviation for the noise in the Doppler shift induced by the thermal noise is the same as that of the thermal noise itself. It is about two orders in magnitude smaller than that of the noise in the Doppler shift.

The \textit{antenna mechanical noise} of the antenna is not previously measured and calculated, thus we cannot estimate its Allan deviation. Here, we will borrow from the experience of the Cassini mission, though the antenna mechanical noise is not the same. The antenna mechanical noise is a random process which may have the positive correlation at the two-way light time(Armstrong et al., 2003). But from the autocorrelation of the residual as shown in Fig. \ref{Fig-omc532023202_56643_vel_autocorr_II_9sInterval}, there is no explicitly positive correlation in the residual, thus we may expect from the Cassini experiment(Asmar et al., 2005) that the Allan deviation of the Doppler shift noise induced by the antenna mechanical noise is smaller than that of the noise in Doppler shift.

For the \textit{transponder noise} and the \textit{ground electronics noise}, they are also not tested. From the analysis of these noises in the Cassini experiment(Asmar et al., 2005), the Allan deviation of the transponder noise may be smaller than $1\times 10^{-13}$. If we assume the same form of the power spectrum $S_{groundelec} \propto f^2$ of the ground electronics noise as that in the Cassini mission, then the Allan deviation of the ground electrical noise is smaller than $3\times10^{-14}$ when $\tau = 9$s. Thus we expect that the ground electronics noise of the Chang'e 3 mission will not contribute significantly to the noise in the Doppler shift.

The \textit{unmodeled mechanical motion of the lander} can also give rise to noise in the Doppler tracking data. The mechanical motion of the lander is mainly generated by the lunar seismic shaking and the solid-body tide on the Moon due to the attraction of the Earth. For the lunar seismic shaking, from the analysis of the events collected from the seismometers of the Apollo program, it was predicted that a ground motion of magnitude larger than 22.5 nm may occur at most once in one year(Mendell, 1998). The time-varying tidal displacements mainly contains a constant term and two dominant periodic terms with periods 27.55 d and 27.21 d, whose amplitudes are all smaller than 0.1 m(Williams et al., 1996). Therefore the noise due to the unmodelled motion of the lander in the Doppler shift is very small. Its Allan deviation is estimated to be smaller than $1\times 10^{-16}$.

Among the propagation noises, the \textit{tropospheric delay noise} in the Doppler tracking is the most important. At microwave frequencies, tropospheric refractive index fluctuations are non-dispersive and dominated by the water vapor fluctuation. Since the data used was obtained on 17 December, 2013, which was winter time at the Jiamusi station, the elevation angles were larger than 20 degrees, the diameter of the antenna was 65m and the Sun-Earth-Moon angle was nearly 180 degrees, thus the Allan deviation $\sigma_{y}(\tau)$ for the Doppler shift noise induced by the troposphere is estimated to be smaller than $5.0\times10^{-14}$(Linfield, 1998) when $\tau = 9$s.

The \textit{ionospheric delay noise} is another source of propagation noise. The ionospheric correction is small in the X-band ($\approx8.47$ GHz ), which is of the order of 10 $\mu m/s$ in magnitude. Since the power spectral density of the ionospheric noises is of the form $S_y(f) \propto f^{-2/3}$(Asmar et al., 2005), and the Sun-Earth-Moon angle is nearly 180 degrees, we may estimate that the Allan deviation $\sigma_{y}(\tau)$ for the Doppler shift noise due to the ionosphere is smaller than $9.1\times 10^{-15}$ with $\tau = 9$s. Thus it is of one to two orders in magnitude smaller than the Allan deviation of the noise in the Doppler shift.

\section{A feasibility study of using the future Chang'e 4 to constrain the SBGWs around 0.01 Hz}

Chang'e 4 was originally built as a backup to the Chang'e 3 mission. After the successful landing of the Chang'e 3 mission on the lunar surface, Chang'e 4 is redefined to land on the far side of the Moon and due to be launched by the end of 2018. To maintain communication between the lander and the ground station, a tracking and data relay satellite (TDRS) will be launched and located at the Earth-Moon L2 point.

\begin{figure}[h]
\begin{center}
\begin{tabular}{c}
\includegraphics[width=0.7\textwidth]{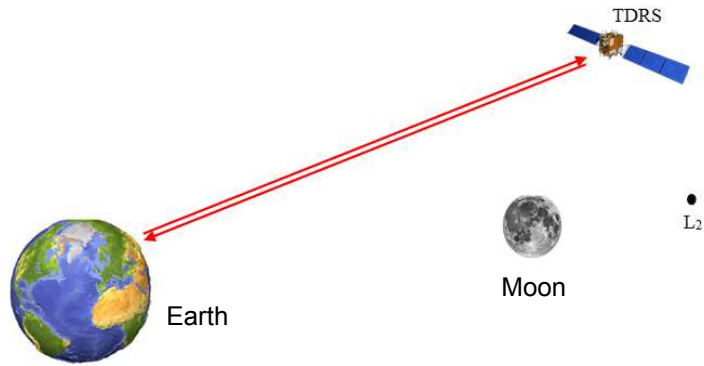}
\end{tabular}
\caption{\label{Fig-CE-4_Scheme_I}The station-TDRS-station link. The radio signal is transmitted from the station to the TDRS and then phase coherently sent back to the station.}
\end{center}
\end{figure}

\begin{figure}[h]
\begin{center}
\begin{tabular}{c}
\includegraphics[width=0.7\textwidth]{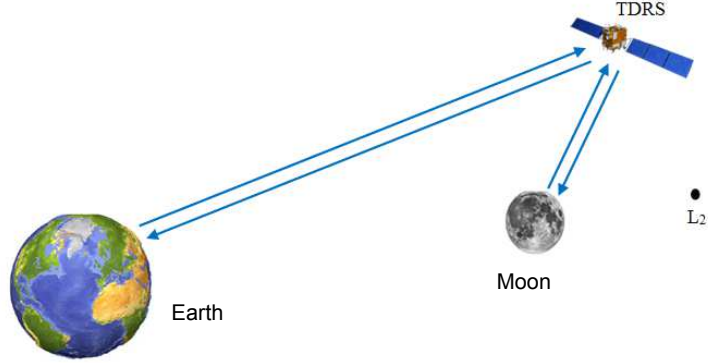}
\end{tabular}
\caption{\label{Fig-CE-4_link_II}The station-TDRS-lander-TDRS-station link. The radio signal is transmitted from the station to the TDRS, then it is phase coherently transmitted from the TDRS to the lander. At the lander, the signal is phase coherently sent back to the TDRS and then sent back to the station.}
\end{center}
\end{figure}

With this mission design, the Doppler tracking of two possible coherent radio links, with frequency standard referenced to a very stable hydrogen clock on ground, will be considered in what follows. The first link is the station-TDRS-station link, as shown in Fig. \ref{Fig-CE-4_Scheme_I}. In this link, the radio signal is transmitted from the station to the TDRS and then phase coherently sent back to the station. The second link is the station-TDRS-lander-TDRS-station link, as shown in Fig. \ref{Fig-CE-4_link_II}. In this link, the radio signal is transmitted from the station to the TDRS, then it is phase coherently transmitted from the TDRS to the lander. At the lander, the signal is phase coherently sent back to the TDRS and then sent back to the station.

From each link, with the upgrade of the radio measurement system of the deep space network, especially with higher precision ultra stable oscillators, the Doppler tracking data with higher precision is expected to be available. Given the upper bound on the SBGWs within the frequency band from 1 mHz to 0.05 Hz attained by the Chang'e 3 mission, it is likely that the Doppler tracking data of the Chang'e 4 mission may yield a better upper bound on the SBGWs in the frequency band from 1 mHz to 0.1 Hz.

\subsection{The first scheme to constrain the SBGWs }
In this scheme, the Doppler tracking data is obtained from the first link, which is denoted by $\dot{\rho}_E^{2w}$. The corresponding baseline is the distance between the station and the TDRS, as shown in Fig. \ref{Fig-CE-4_Scheme_I}, which is about $4.2\times 10^5$ km. The corresponding round-trip time, denoted by $T_2^E$, of the radio signal along the first link is about 2.80 s. After subtracting the theoretical values from the observations, the residual $y_I$ in the Doppler shift $\dot{\rho}_E^{2w}/c$ ( $c$ is the speed of light) may be modelled as
\begin{eqnarray}\label{NoiseModelCE4_I}
  y_I(t) &=& y_{2,E}^{gw}(t) + y^{trop}(t) + y^{trop}(t - T_2^E) + y^{iono}(t)+y^{iono}(t-T_2^E)  \nonumber\\
   &&+ y^{ant}(t) + y^{ant}(t-T_2^E) + y^{FTS}(t) - y^{FTS}(t-T_2^E)            \nonumber\\
   &&+ y^{TDRS}(t-T_2^E/2) + y^{ther}(t) + y^{groundelec}(t),
\end{eqnarray}
where $y_{2,E}^{gw}$ is the contribution of the SBGWs to the Doppler variability, the other terms are the main noise sources of variability in the Doppler shift. The meanings of the noise terms $y^{trop}$, $y^{iono}$, $y^{ant}$, $y^{FTS}(t)$, $y^{ther}$ and $y^{groundelec}$ are the same as those given in Eq.(\ref{CE-3NoiseTransfer}). The noise $y^{TDRS}$ represents the transponder noise in the TDRS. In the spectral domain, the power spectrum $S_{y_I}(f)$ of $y_I$ is given by
\begin{eqnarray}\label{NoiseModelCE4PSD_I}
  S_{y_I}(f) &=& S_{y_{2,E}}^{gw}(f) + 4\cos^2(\pi T_2^E f)\left[S_{y}^{trop}(f) + S_{y}^{iono}(f)+ S_{y}^{ant}(f) \right] \nonumber\\
   && + 4\sin^2(\pi T_2^E f)S_{y}^{FTS}(f)+ S_{y}^{TDRS}(f) + S_{y}^{groundelec}(f)+ S_{y}^{ther}(f), \nonumber\\
\end{eqnarray}
where $S_{y_{2,E}}^{gw}$ is the power spectrum of the term $y_{2,E}^{gw}$ due to the SBGWs and the other terms are the corresponding power spectrums of the noises in Eq.(\ref{NoiseModelCE4_I}).

Suppose the power spectrum $S_{y_I}(f)$ of $y_I$ is white in the band from 1 mHz to 0.1 Hz. It may be modelled as $S_{y_I}(f) = \frac{\sigma_{I}^2}{2\pi}$, where $\sigma_{I}$ represents the root mean square of the residual $y_I$. For the power spectrum $S_{y_I}(f)$ of $y_I$ to give an improved upper bound on the SBGWs, the main noises must satisfy certain  requirements, which are easily evaluated from Eq. (\ref{NoiseModelCE4PSD_I}) and listed in the Table \ref{Tab-RequirementOnNoise_I}.
\begin{table}[h]
\begin{center}
\begin{tabular}{|l|l|}
  \hline
Noise             & Requirement on psd    \\
  \cline{1-2}
Troposphere       & $ \leq \frac{\sigma_{I}^2}{8\pi\cos^2(\pi T_2^E f )}$ \\
   \cline{1-2}
Ionosphere        & $\leq \frac{\sigma_{I}^2}{8\pi\cos^2(\pi T_2^E f )}$ \\
   \cline{1-2}
Antenna           & $\leq \frac{\sigma_{I}^2}{8\pi\cos^2(\pi T_2^E f )}$ \\
   \cline{1-2}
Clock             & $\leq \frac{\sigma_{I}^2}{8\pi\sin^2(\pi T_2^E f )}$ \\
   \cline{1-2}
TDRS              & $\leq \frac{\sigma_{I}^2}{2\pi}$ \\
   \cline{1-2}
Ground electric   & $\leq \frac{\sigma_{I}^2}{2\pi}$ \\
   \cline{1-2}
Thermal           & $\leq \frac{\sigma_{I}^2}{2\pi}$ \\
  \hline
\end{tabular}
\caption{Requirements on the power spectral density of the noises in the first scheme. }
\label{Tab-RequirementOnNoise_I}
\end{center}
\end{table}

\subsection{The second scheme to constrain the SBGWs }

In Chang'e 4, the simultaneously obtained two coherent radio links will share the same uplink carrier wave, and for the downlink waves, they will pass the same space path between the relay satellite and the ground tracking antenna, as well as the receiver, the low noise amplifier, and the Doppler counter. Thus it may be possible to reduce or cancel out many common noise fluctuations by the combination of the two kinds of Doppler tracking.

Let $\dot{\rho}^{4w}(t)$ denote the range rate given from the second link. The residual $y_4$ in the Doppler shift $\dot{\rho}^{4w}(t)/c$ may be modelled as
\begin{eqnarray}\label{NoiseModel4wRR}
  y_4(t) &=& y_4^{gw}(t) + y^{trop}(t) + y^{trop}(t - T_2^E-T_2^M) + y^{iono}(t)+y^{iono}(t-T_2^E-T_2^M)  \nonumber\\
   &&+ y^{ant}(t)+y^{ant}(t-T_2^E-T_2^M)+y^{FTS}(t)-y^{FTS}(t-T_2^E-T_2^M)      \nonumber\\
   && + y^{TDRS}(t-T_2^E/2) + y^{TDRS}(t-T_2^E/2-T_2^M)        \nonumber\\
   &&+ y^{ldr}(t-T_2^E/2-T_2^M/2)+y^{groundelec}(t) + y^{ther}(t),
\end{eqnarray}
where $T_2^M$ is the round-trip time of the signal from the TDRS to the lander and then return back to the TDRS, the $y_4^{gw}$ is the contribution of the SBGWs to the Doppler variability, $y^{ldr}$ is the noise in the Doppler shift due to the transponder noise of the lander. The meanings of the other terms are the same as those in Eq. (\ref{NoiseModelCE4_I}).

From the noise analysis of Chang'e 3, we know that the main noises are the clock noise, the tropospheric noise and the antenna mechanical noise. It follows from Eq. (\ref{NoiseModelCE4_I}) and Eq. (\ref{NoiseModel4wRR}) that to reduce or cancel out their influences, we construct a new observable $O_2(t)$ as $O_2(t)=\dot{\rho}^{4w}(t)-\dot{\rho}_E^{2w}(t)$. If we denote $y_{II}(t)$ as the residual in the Doppler shift $O_2(t)/c$, then it equals to the differential of $y_4$ and $y_I$. The residual $y_{II}(t)$ may be modelled from Eq. (\ref{NoiseModelCE4_I}) and Eq. (\ref{NoiseModel4wRR}) as
\begin{eqnarray}\label{NoiseModelCE4_II}
  y_{II}(t) &=& y_4^{gw}(t) - y_{2,E}^{gw}(t) - y^{trop}(t - T_2^E) + y^{trop}(t - T_2^E-T_2^M) \nonumber \\
  && - y^{iono}(t - T_2^E)+y^{iono}(t-T_2^E-T_2^M) - y^{ant}(t - T_2^E) \nonumber\\
  && + y^{ant}(t-T_2^E-T_2^M)+y^{FTS}(t - T_2^E)-y^{FTS}(t-T_2^E-T_2^M)      \nonumber\\
  && + y^{TDRS}(t-T_2^E/2-T_2^M) + y^{ldr}(t-T_2^E/2-T_2^M/2).
\end{eqnarray}
Obviously, the influence of the SBGWs to the Doppler shift is $y_4^{gw}(t) - y_{2,E}^{gw}(t)$, which is in fact equivalent to the influence of the SBGWs to the Doppler shift given from the two-way range rate between the lander and the TDRS, thus we may denote by $y_{2,M}^{gw}$ the $y_4^{gw}(t) - y_{2,E}^{gw}(t)$. In this scheme, the distance between the lander and the TDRS is about $6.5\times 10^4$ km. So the round-trip time $T_2^M$ is about 0.43 s.

The power spectrum $S_{y_{II}}$ of $y_{II}$ may be evaluated as
\begin{eqnarray}\label{NoiseModelCE4PSD_II}
  S_{y_{II}}(f) &=& S_{y_{2,M}}^{gw}(f) + 4\sin^2(\pi T_2^M f)\left[S_{y}^{trop}(f) + S_{y}^{iono}(f)+ S_{y}^{ant}(f) \right] \nonumber\\
   && + 4\sin^2(\pi T_2^M f)S_{y}^{FTS}(f)+ S_{y}^{TDRS}(f) + S_{y}^{ldr}(f),
\end{eqnarray}
where $S_{y_{2,M}}^{gw}(f)$ is the power spectrum of $y_{2,M}^{gw}$ and $S_{y}^{ldr}(f)$ is the power spectrum of $y^{ldr}$.

Assume that the power spectrum $S_{y_{II}}(f)$ of the residual $y_{II}(t)$ is constant in the frequency band from 1 mHz to 0.1 Hz. It may be written as $S_{y_{II}}(f) = \frac{\sigma_{II}^2}{2\pi}$, where $\sigma_{II}$ is the root mean square of the residual $y_{II}$. The requirements on the main noises are obtained from Eq. (\ref{NoiseModelCE4PSD_II}), which are listed in the Table \ref{Tab-RequirementOnNoise_II}.

\begin{table}[h]
\begin{center}
\begin{tabular}{|l|l|}
  \hline
Noise             & Requirement on psd    \\
  \cline{1-2}
Troposphere       & $\leq \frac{\sigma_{II}^2}{8\pi\sin^2(\pi T_2^M f)}$ \\
   \cline{1-2}
Ionosphere        & $\leq \frac{\sigma_{II}^2}{8\pi\sin^2(\pi T_2^M f)}$ \\
   \cline{1-2}
Antenna           & $\leq \frac{\sigma_{II}^2}{8\pi\sin^2(\pi T_2^M f)}$ \\
   \cline{1-2}
Clock             & $\leq \frac{\sigma_{II}^2}{8\pi\sin^2(\pi T_2^M f)}$ \\
   \cline{1-2}
TDRS              & $\leq \frac{\sigma_{II}^2}{2\pi}$ \\
   \cline{1-2}
Lander            & $\leq \frac{\sigma_{II}^2}{2\pi}$ \\
  \hline
\end{tabular}
\caption{Requirements on the power spectral density of the noises in the second scheme. }
\label{Tab-RequirementOnNoise_II}
\end{center}
\end{table}

\subsection{Comparisons of two schemes}

\begin{figure}[h]
\begin{center}
\begin{tabular}{c}
\includegraphics[width=0.7\textwidth]{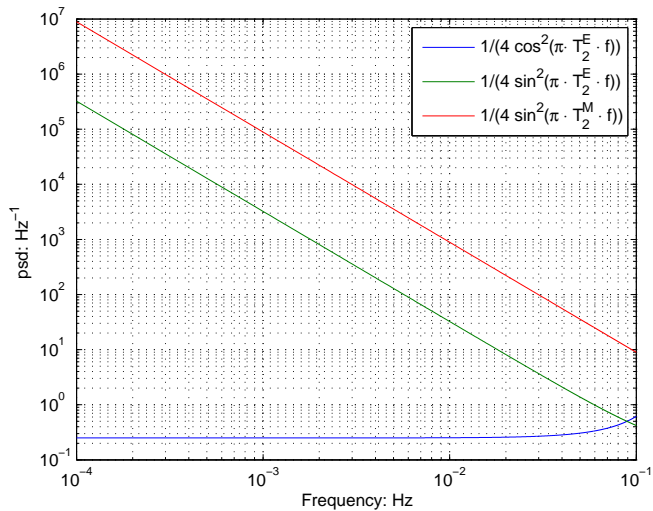}
\end{tabular}
\caption{\label{Fig-CE4_specialFun}The plot for the functions $1/(4\cos^2(\pi T_2^E f))$, $1/(4\sin^2(\pi T_2^E f))$ and $1/(4\sin^2(\pi T_2^M f))$.}
\end{center}
\end{figure}

It follows from the Tables \ref{Tab-RequirementOnNoise_I} and \ref{Tab-RequirementOnNoise_II} that these two schemes have different noise requirements. In the first scheme, there is a common noise requirement on the tropospheric noise, the ionospheric noise and the antenna mechanical noise, namely, their power spectrums should be smaller than $\frac{\sigma_{I}^2}{8\pi\cos^2(\pi T_2^E f )}$. The power spectrum of the clock should be smaller than $\frac{\sigma_{I}^2}{8\pi\sin^2(\pi T_2^E f )}$. In the second scheme, the noise requirements on the tropospheric noise, the ionospheric noise, the antenna mechanical noise and the clock noise are identical. Their power spectrums should be smaller than $\frac{\sigma_{II}^2}{8\pi\sin^2(\pi T_2^M f)}$. For comparison, we show in Fig. \ref{Fig-CE4_specialFun} the functions $1/(4\cos^2(\pi T_2^E f))$, $1/(4\sin^2(\pi T_2^E f))$ and $1/(4\sin^2(\pi T_2^M f))$. We find that when $\sigma_{I} = \sigma_{II}$, the requirements on the tropospheric noise, the ionospheric noise and the antenna mechanical noise in the first scheme are about 1.5 orders in magnitude higher than those in the second scheme. Further the requirement on the clock noise in the first scheme is also much higher than that in the second scheme. For example, the requirement in the first scheme is about 3.5 orders in magnitude higher than that in the second scheme at 0.01 Hz.

From the noise analysis of Chang'e 3, the main noises limiting the improvement of the measurement accuracy are the propagation noises and the antenna mechanical noise. Thus the second scheme is better than the first scheme. Further, in the second scheme, we do not have to consider the ground electronic noise and thermal noise. Thus, the second scheme is recommended for future Doppler tracking of Chang'e 4 to constrain the SBGWs. In the following, we will further elaborate on certain practical aspects of the main noises when we try to implement the measurement scheme.

\begin{figure}[h]
\begin{center}
\begin{tabular}{c}
\includegraphics[width=0.7\textwidth]{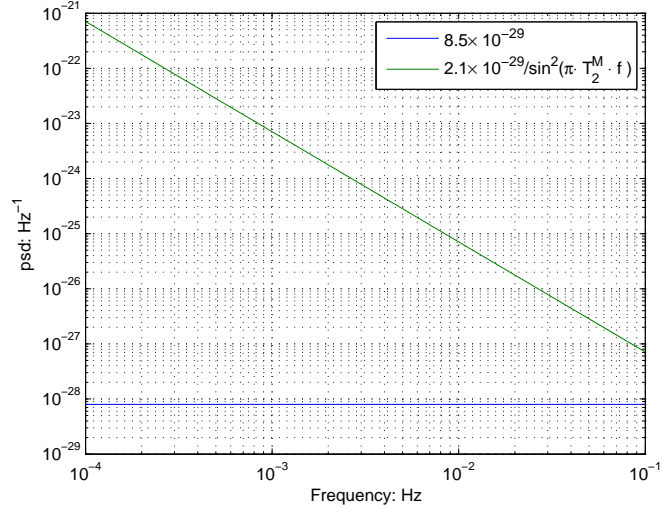}
\end{tabular}
\caption{\label{Fig-CE4_specialFun_II_NoiseReq}The plot for the functions $2.1\times10^{-29}/\sin^2(\pi T_2^M f )$ and $8.5\times10^{-29}$.}
\end{center}
\end{figure}

In the Chang'e 4 mission, the measurement accuracy of the range rate is expected to reach about 15$\mu$m/s with the sampling time to be 1 s. Since we are interested in constraining the SBGWs in the frequency band from 1 mHz to 0.1 Hz, the original Doppler tracking can be smoothed to a new time series with the sampling time to be 5 s. Thus in principle the root mean squares $\sigma_{I}$ and $\sigma_{II}$ are about $2.3\times 10^{-14}$. It then follows from the Table \ref{Tab-RequirementOnNoise_II} that the explicitly requirements on the main noises are listed in the Table \ref{Tab-RequirementOnNoise_II_15}. For clarity, we plot the functions $2.1\times10^{-29}/\sin^2(\pi T_2^M f )$ and $8.5\times10^{-29}$ in the Fig. \ref{Fig-CE4_specialFun_II_NoiseReq}.
\begin{table}[h]
\begin{center}
\begin{tabular}{|l|l|}
  \hline
Noise             & Requirement on psd    \\
  \cline{1-2}
Troposphere       & $\leq 2.1\times10^{-29}/\sin^2(\pi T_2^M f)$ \\
   \cline{1-2}
Ionosphere        & $\leq 2.1\times10^{-29}/\sin^2(\pi T_2^M f)$ \\
   \cline{1-2}
Antenna           & $\leq 2.1\times10^{-29}/\sin^2(\pi T_2^M f)$ \\
   \cline{1-2}
Clock             & $\leq 2.1\times10^{-29}/\sin^2(\pi T_2^M f)$ \\
   \cline{1-2}
TDRS              & $\leq 8.5\times10^{-29}$ \\
   \cline{1-2}
Lander            & $\leq 8.5\times10^{-29}$ \\
  \hline
\end{tabular}
\caption{Requirements on the power spectral density of the noises in the second scheme when $\sigma_{II} = 2.3\times 10^{-14}$. }
\label{Tab-RequirementOnNoise_II_15}
\end{center}
\end{table}

For the tropospheric noise, its power spectral density is(see Linfield, 1998)
\begin{eqnarray}\label{psd_Troposphere}
  S_{y}^{trop}(f) & = & 1.4\times10^{-27}f^{-2/5} \textrm{Hz}^{-1}, \quad 10^{-5}\leq f \leq 10^{-2} \textrm{Hz},\\
                  & = & 2.2\times10^{-30}f^{-3}   \textrm{Hz}^{-1}, \quad 10^{-2} \leq f \leq 1 \textrm{Hz}.
\end{eqnarray}
It is smaller than that requirement listed in the Table \ref{Tab-RequirementOnNoise_II_15} when the frequency is smaller than 0.01 Hz. But it is about 0.5 to 1 order in magnitude larger than the requirement from 0.01 to 1 Hz. Since it is well known that the tropospheric noise depends on the elevation angle, the season, time of day, and weather conditions, it is likely we may lower the tropospheric noise to reach the requirement.

For the ionospheric noise, for the X-band($\approx$ 8.47GHz ), its power spectrum $S_{y}^{iono}(f)$ is approximated by $7.79\times10^{-29}f^{-2/3} \textrm{Hz}^{-1}$(Tinto, 2002). Thus it is smaller than that in the Table \ref{Tab-RequirementOnNoise_II_15}.

For the ground master clock and the frequency and timing distribution, if we use a clock with the one-sided power spectral density given as(Tinto et al., 2009)
\begin{eqnarray}
  S_{y}^{FTS}(f) &=& 6.2\times[10^{-28}f + 10^{-33}f^{-1}]  + 1.3\times10^{-28}f^2 \quad \textrm{Hz}^{-1},
\end{eqnarray}
then the clock noise satisfies the requirement listed in the Table \ref{Tab-RequirementOnNoise_II_15}.

As far as antenna mechanical noise is considered, if we convert from the required power spectrum of the antenna mechanical noise to the Allan deviation $\sigma_{ant}(\tau)$, we have that $\sigma_{ant} \approx 3.2\times10^{-13}$ when $\tau = 5$s. From the experience of Cassini, it may be reached if the observation is under some favorable conditions(Asmar et al., 2005), which we will study carefully in future.

For the TDRS and the lander, the main noises are the transponder noise(Asmar et al., 2005). In the Cassini spacecraft, the one-sided power spectral density $S_{y}^{TR}(f)$ is(Riley et al., 1990)
\begin{eqnarray}
  S_{y}^{TR}(f) &=& 1.6\times10^{-26}f \quad \textrm{Hz}^{-1} .
\end{eqnarray}
This power spectrum is $1.6\times10^{-29}$ at about 1 mHz and then increases linearly to $1.6\times10^{-27}$ at 0.1 Hz which is about 1.2 orders in magnitude higher than the required $8.5\times10^{-29}$. Thus it is possible to improve the transponder noise on the TDRS and the lander.

\begin{figure}[h]
\begin{center}
\begin{tabular}{c}
\includegraphics[width=0.4\textwidth]{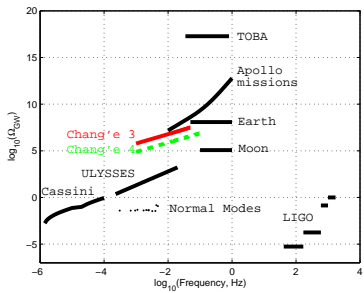}
\end{tabular}
\caption{\label{Fig-CE4_SBGW_Omega_comparison_VII_10s_EandW}The expected upper bound on SBGWs from Chang'e 4 (green dash line). The other constraints are listed here for comparison. These constrains include the results obtained from the Cassini spacecraft(Armstrong et al., 2003), the ULYSSES spacecraft(Bertotti et al., 1995), the normal modes of the Earth(Coughlin et al., 2014a), the Apollo missions(Aoyama et al., 2014), the Earth's seismic data(Coughlin et al., 2014b), the Lunar seismic data(Coughlin et al., 2014c), the torsion-bar antenna(Shoda et al., 2013), the LIGO mission(Aasi et al., 2014)(corresponds to four lines) and the Chang'e 3 mission (red line).}
\end{center}
\end{figure}

Subject to the requirements on the noises in the second scheme, it is expected that the best measurement accuracy of the data is about 15$\mu$m/s, then the upper bound on the energy density $\Omega_{gw}$ of SBGWs in the band from 1 mHz to 0.1 Hz is calculated and shown in Fig. \ref{Fig-CE4_SBGW_Omega_comparison_VII_10s_EandW}. It indicates that the upper bound on $\Omega_{gw}$ will be improved nearly one order in a wider band when compared with that of Chang'e 3. Further, it will be improved by nearly 1.3 orders at about 0.1 Hz when compared with those from the Earth's seismic data.

\section{Conclusion}

A detailed analysis has been presented on the range rate data of the Chang'e 3 lunar mission. Apart from giving an improved upper bound on the SBGWs in the narrow frequency band of 0.02 to 0.05 Hz, the feasibility of improving on the upper bound in the upcoming Chang'e 4 mission is also discussed. By making use of the differential coherent Doppler measurement in Chang'e 4, the effects due to the interplanetary plasma, ionosphere, troposphere, and the effects due to the ground instruments and clock can be removed considerably. The upper bound on the SBGWs may then be improved by nearly one order in a wider band when compared with that of Chang'e 3. With the promise in science for future lunar as well as deep space programs in China, we hope our work constitutes a modest beginning on this front as far as experimental tests of general relativity are concerned.

\section*{Acknowledgment}
The present research is supported by the Strategic Priority Research Program of the Chinese Academy of Sciences��Grant No.XDB23030100, National Natural Science Foundation of China (project number 11173005, 11171329, 61304233 and 41590851), and also in part by the National Basic Research Program of China under Grant 2015CB857101.





\end{document}